\documentclass[aps,pre,twocolumn,groupedaddress,longbibliography]{revtex4-1}

\usepackage{amsmath}
\usepackage{graphicx}
\usepackage{color}
\usepackage{epstopdf}
\usepackage{upgreek}
\usepackage{array}
\usepackage{multirow}
\usepackage{tabularx}
\usepackage{CJK}
\usepackage{lineno}
\usepackage{setspace}
\usepackage{xcolor}
\pagecolor{white}

\begin{document}

\singlespacing

\newcommand{\red}{\textcolor{red}}
\newcommand{\green}{\textcolor{green}}
\newcommand{\blue}{\textcolor{blue}}
\newcolumntype{M}[1]{>{\centering\arraybackslash}m{#1}}

\begin{CJK*}{UTF8}{}
\title{Experimental determination of the propulsion matrix of the body of helical \textit{Magnetospirillum magneticum} cells}

\author{Liu Yu, Lucas Le Nagard, Solomon Barkley, Lauren Smith, C\'ecile Fradin}
\affiliation{Department of Physics and Astronomy, McMaster University, 1280 Main St. W, Hamilton, ON L8S4M1, Canada}

\begin{abstract}
Helical-shaped magnetotactic bacteria provide a rare opportunity to precisely measure both the translational and rotational friction coefficients of micron-sized chiral particles. The possibility to align these cells with a uniform magnetic field allows to clearly separate diffusion along and perpendicular to their longitudinal axis. Meanwhile, their corkscrew shape allows detecting rotations around their longitudinal axis, after which orientation correlation analysis can be used to retrieve rotational diffusion coefficients in the two principal directions. Using light microscopy, we measured the four principal friction coefficients of deflagellated \textit{Magnetospirillum magneticum} AMB-1 cells, and compared our results to that expected for cylinders of comparable size. We show that for rotational motions, the overall dimensions of the cell body are what matters most, while the exact body shape influences translational motions. To obtain a full characterization of the friction matrix of these elongated chiral particles, we also quantified the coupling between the rotation around and translation along the longitudinal axis of the cell. Our results suggest that for this bacterial species cell body rotation could significantly contribute to cellular propulsion. 

\end{abstract}


\maketitle
\end{CJK*}


\section{Introduction}

The swimming of microorganisms takes place at low Reynolds number, where viscous forces play a dominant role \cite{Gray1955,Lighthill1976,purcell1977life,Berg1993}. As such, modeling the motion of these organisms requires a precise knowledge of their friction coefficients. In the case of flagellated bacteria, swimming involves opposite rotations of the cell body and flagella, resulting in an overall translation along the propulsion axis \cite{berg1973bacteria,macnab1977bacterial,darnton2007torque}. Meanwhile, the tumbling motion used by many bacteria (notably \textit{Escherichia coli}) to change direction and achieve chemotaxis involves a rotational diffusion of the propulsion axis \cite{macnab1972gradient,berg1972chemotaxis}. And both translation along the propulsion axis and rotations are involved in the "U-turn" motion of magnetotactic bacteria (MTB) submitted to a magnetic field reversal \cite{Esquivel1986,mohammadinejad2021stokesian}. These examples show the importance of determining all the friction coefficients of a particular microorganism (translational and rotational, along and perpendicular to the cell longitudinal axis) in order to fully understand its motility. 

Although friction coefficients can in principle be calculated for bacteria with a cylindrical shape such as \textit{E. coli}, things become more complicated for cells with more asymmetrical shapes. Here we are interested in the magnetotactic species \textit{Magnetospirillum magneticum}, with the characteristic ``corkscrew" shape representative of spirilla. The friction coefficients of spiral bacteria have been approximated by treating the cells as spheres \cite{Esquivel1986}, linear chains of spheres \cite{Bahaj1996}, cylinders \cite{chwang1972locomotion,Nadkarni2013} or prolate spheroids \cite{Reufer2014}. A more accurate model was recently obtained by taking the actual helical shape of the cells into account using finite element analysis \cite{zahn2017measurement}. An experimental approach, involving the construction of macroscopic models of spiral cells, was also use to estimate their friction coefficients \cite{pichel2018magnetic}. None of these strategies, however, accounts for the exact cellular shape, including irregularities and eventual appendages. In addition, none of the above studies delved into the coupling between rotation and translation expected for chiral objects.

Here we propose to experimentally measure the friction coefficients of deflagellated cells of the spirillum \textit{M. magneticum} by recording their translational and rotational diffusion as observed with light microscopy. We take advantage of the asymmetric shape of spirilla, which allows a full determination of a cell's orientation from its projection in the focal plane \cite{constantino2016helical,LeNagard2019}. We also take advantage of the magnetic properties of \textit{M. magneticum}, which allows aligning the average direction of the longitudinal axis of the cells with that of an external magnetic field, and separately measuring transversal and longitudinal friction coefficients. We first show the results of simulations used to determine the best experimental strategy for extraction of the different friction coefficients of cells from their trajectories. We then present measurements of the friction coefficients as a function of cell length and compare them with different available theoretical models. Finally we experimentally quantify for the first time the coupling between rotation around and translation along the cell longitudinal axis.

\section{Methods} \label{methods}

\subsection{Simulations of rotational diffusion} \label{methods:simulations}

\begin{figure}
\includegraphics[width=6 cm]{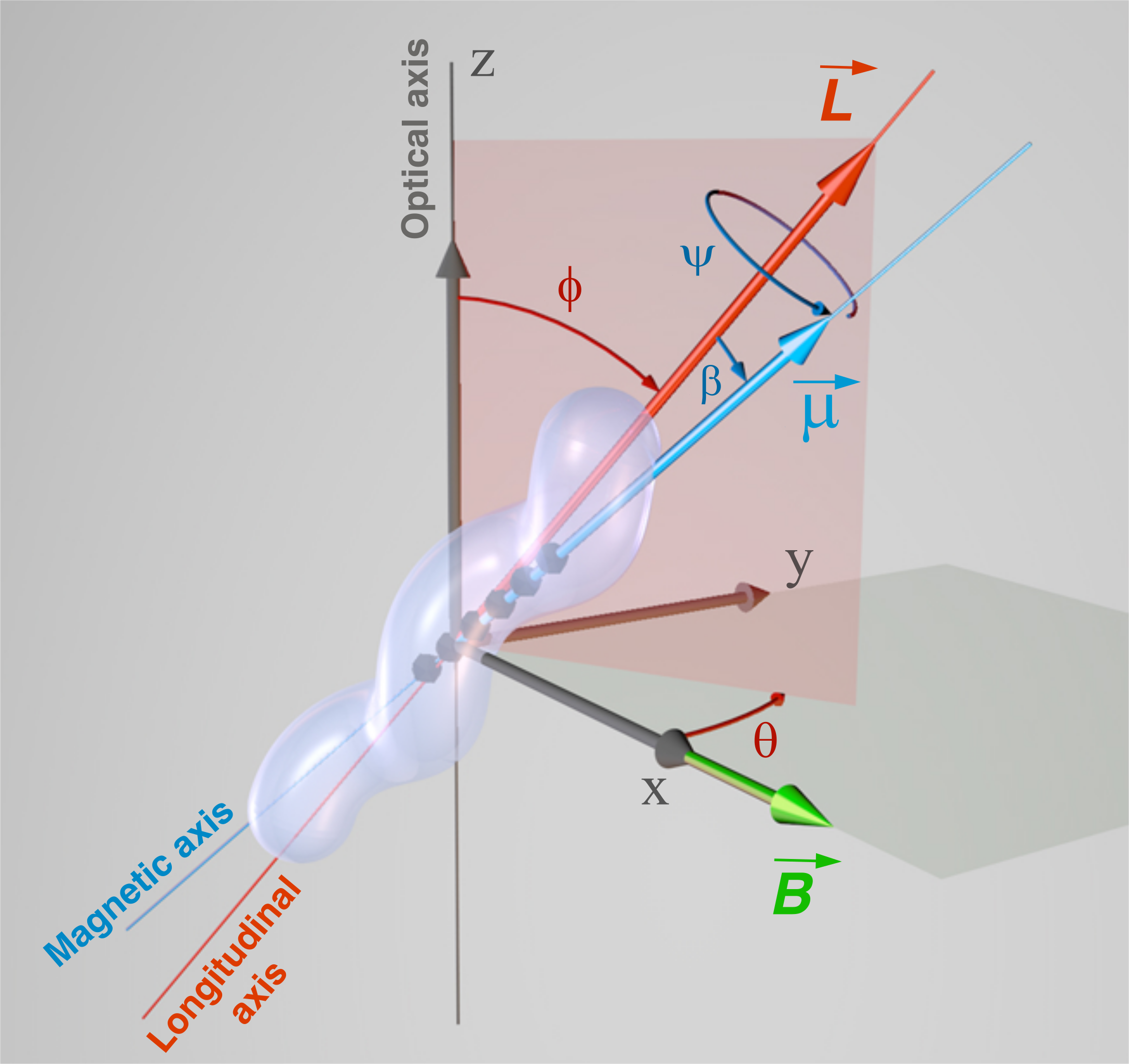}
\caption{Orientation of the cell with respect to the focal plane and external magnetic field ($\vec{B}$). The position of the cell longitudinal axis (defined by $\vec{L}$) is characterized by its inclination $\phi$ from the optical axis and by the angle $\theta$ between its projection in the focal plane and the magnetic field. The direction of the cell magnetic moment ($\vec{\mu}$) is characterized by the angle $\beta$ it makes with $\vec{L}$ and by the rotation $\psi$ of the cell around its longitudinal axis.}
\label{fig:axes}
\end{figure}

Simulations of the rotational diffusion of a magnetotactic cell placed in a uniform magnetic field were performed using Mathematica (Wolfram Research). The cell was assimilated to an elongated rigid body with rotational symmetry around its longitudinal axis ($\vec{L}$), and rotational friction coefficients $f_{r_{\perp}}$ and $f_{r_{\parallel}}$, associated respectively with rotations perpendicular to and around $\vec{L}$. The cell's magnetic moment, $\vec{\mu}$, was placed at a constant inclination ($\beta$) from $\vec{L}$, as illustrated in Fig.~\ref{fig:axes}. The orientation of the cell was updated every $\delta t = 1$~ms, by performing a series of $4$ small rotations. The cell was first allowed to diffuse around its three principal axes of rotation ($\vec{L}$, $\vec{L} \times \vec{z}$ and $\vec{L} \times (\vec{L} \times \vec{z})$). For the first rotation, the angular displacement was drawn from a Gaussian distribution with variance $2 D_{r_\parallel} \delta t$, where $D_{r_\parallel}=kT/f_{r_\parallel}$ is the axial rotational diffusion coefficient. For the last two rotations, the transversal rotational diffusion coefficient $D_{r_{\perp}}=kT/f_{r_{\perp}}$ was used. In the presence of an external magnetic field ($\vec{B}$) an additional rotation was added to account for the magnetic torque. The cell as a whole (\textit{i.e.} both $\vec{L}$ and $\vec{\mu}$) was rotated around $\vec{\mu} \times \vec{B}$, by an angle $| \vec{\mu} \times \vec{B} | \delta t / f_{r_\perp} $ (using $f_{r_\perp}$ as the friction coefficient, which should be a very good approximation as long as $\beta$ is small). At each step, $\theta$ and $\psi$ were calculated from $\vec{L}$ and $\vec{\mu}$: The apparent orientation of the cell in the focal plane ($\theta$) was obtained by projecting $\vec{L}$ onto the $(\vec{x}, \vec{y})$ plane. The apparent rotation around its longitudinal axis ($\psi$) was calculated as the angle between the $(\vec{L}, \vec{z})$ plane (red plane in Fig.~\ref{fig:axes}) and the vector $\vec{\mu} - (\vec{\mu} \cdot \vec{L}/L^2) \vec{L}$.
Simulations were typically run for $2000$ steps ($2$ s) at $T = 300$ K, using physical parameters representative of those expected for \textit{M. magneticum} cells: $D_{r_{\perp}} = 0.1$ s$^{-1}$, $D_{r_{\parallel}} = 0.01 - 0.5$ s$^{-1}$ and $\mu = 0.5 \times 10^{-15}$ A$\cdot$m$^2$ \cite{LeNagard2019}.
 
\subsection{Cell culture}

Cells of \textit{M. magneticum} strain AMB-1 (obtained from ATCC, 700264) were grown according to the protocol detailed in Ref.~\cite{LeNagard2018}. Briefly, cells were grown at 30 $^{\circ}$C in $60$ mL of growth medium, containing trace mineral supplements, KH$_{2}$PO$_{4}$, MgSO$_{4}$$\cdot$7H$_{2}$O, HEPES, NaNO$_{3}$, yeast extract, soy bean peptone (BD Bacto Soytone), potassium lactate and Fe(III) citrate (pH 7.0), inside $125$ mL sealed glass bottles. Any O$_2$ in the headspace of the bottle and dissolved in the medium was removed by bubbling N$_{2}$ in the headspace and in the solution. The medium was then autoclaved to ensure sterility. Right before inoculation, $1$ mL of O$_{2}$ was added to the headspace ($65$ mL) to reach a $1.5 \%$ O$_{2}$ mircoaerobic environment ideal for the growth of AMB-1 cells with strong magnetic properties. When needed in order to obtain data for longer cells, $10~\upmu$g/mL of cephalexin (Sigma-Aldrich), an antibiotic which can block cell division \cite{pogliano1997inactivation,Katzmann2011magnetosome}, was added to the growth medium two days after inoculation.

\subsection{Cell imaging}

Cells were harvested $3$ to $5$ days after inoculation, then killed and deflagellated by heating at 60 $^{\circ}$C for $15$ min. After cooling down to room temperature, the bacteria suspension was diluted 50-fold in fresh medium to achieve an ideal concentration for single cell observation. The diluted solution was then injected into a home-built sample chamber consisting of a glass slide and a microscope coverslip separated by two melted parafilm strips. The chamber was sealed with vacuum grease or transparent nail polish to avoid evaporation and flow. Movies of cells undergoing translational and rotational diffusion were then immediately captured at 100 frames per second with a fast CCD camera (AVT Prosilica GE) mounted on a phase-contrast microscope (Nikon Eclipse E200-LED), with either a $40 \times$ ($0.65$ NA, pixel size 0.18 $\upmu$m) or a $100 \times$ ($1.25$ NA, pixel size 0.07 $\upmu$m) objective. The effective viscosity of the medium is larger than in the bulk when bacteria are close to a surface \cite{giacche2010hydrodynamic}. To avoid this issue, when studying cell diffusion we only imaged cells that were at least $20~\upmu$m away from the coverslip or glass slide. The stage of the microscope was modified with a pair of custom-made electromagnetic coils \cite{LeNagard2019}, so that constant uniform magnetic fields up to $1.7$ mT could be applied parallel to the focal plane by circulating a current through the coils. The average movie duration was $17$ s, with no movie shorter than $4$ s. When required, cells were immobilized in a hydrogel prepared from $17$ mg/mL agar in deionized water \cite{wong2010agar}. The mixture was microwaved for several seconds until the agar was completely dissolved and injected into a warm sample chamber ($\approx 350~\upmu$m  thick). The chamber was then immediately submerged into a fresh MTB culture, and a magnetic bar was used to impart a vertical orientation to the cells as the gel solidified.

\subsection{Cell tracking} \label{tracking}

\begin{figure}
\includegraphics[width = 7.5cm]{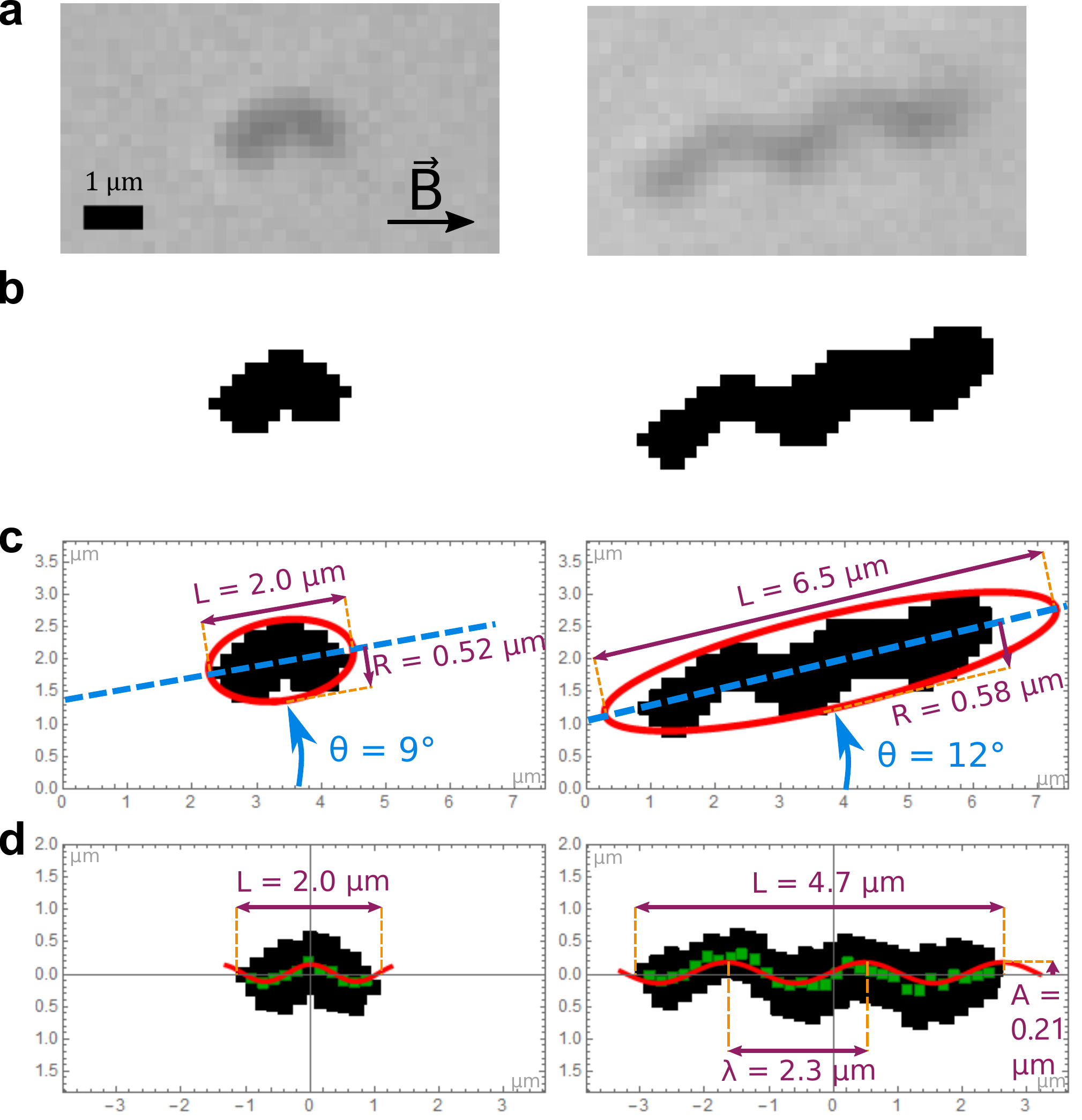}
\caption{Illustration of the image analysis process for a short cell (left panels) and a long cell (right panels). (a) Phase microscopy image of the cell, (b) binarized image, (c) result of the ellipse fit and (d) result of the sine fit after rotation of the cell to an horizontal position.}
\label{fig:cellfit}
\end{figure}

Cells were tracked using the algorithm illustrated in Fig.~\ref{fig:cellfit}. Images of individual cells were first binarized using ImageJ \cite{schneider2012nih}, resulting in a connected cloud of points representing the cell in each frame of the movie (Fig.~\ref{fig:cellfit}b). The position of the cell was tracked by finding the center of mass of this cloud of points. The apparent orientation of the cell was determined in two different ways using a code written in Mathematica. First, the cell was fit with an ellipse to obtain an estimate of the cell apparent orientation, $\theta$, its length, $L$, and diameter, $R$ (Fig.~\ref{fig:cellfit}c). While this method is fast and robust, it may not always capture the exact orientation of helical AMB-1 cells properly, thus a more refined fit was then performed to take into account the sinusoidal shape of the cell body projected in the focal plane, as first described in Ref.~\cite{LeNagard2019} and as illustrated in Fig.~\ref{fig:cellfit}d. The approximate orientation of the cell was first quickly determined using a linear fit, then the cell was rotated so as to lay approximately horizontal. Points were binned vertically to obtain a new series of points (green points in Fig.~\ref{fig:cellfit}d) considered as the cell backbone, which was then fit with a sine function, $A \sin(2 \pi x / \lambda + \psi)$, returning the amplitude ($A$) and wavelength ($\lambda$) of the cell helical backbone, as well as a phase ($\psi$) giving a direct representation of the rotation of the cell around its long axis (as long as the cell lays in the focal plane, or close to it). To further refine the determination of the cell apparent orientation, the horizontal binarized cell image was rotated from $-8.5^{\circ}$ to $+8.5^{\circ}$ in 0.5$^{\circ}$ increments and the backbone determination and sine fit repeated at each step. The results of the fit with the least chi-square were saved. To speed up the image analysis process, this full procedure was only performed for the first $100$ frames. In the rest of the frames, the values of $A$ and $\lambda$ were fixed to the average values obtained from the first $100$ fits, and only the parameters $\theta$, $\psi$ and $L$ were determined.
The cell radius $r$ (Fig.~\ref{fig:geometry}) was measured manually using ImageJ from images obtained at 100$\times$ magnification. When comparing the results of the measurement of the apparent orientation of the cells in the focal plane with either the sine fit or the elliptical fit for the same images, we found that there was on average a $3.2^\circ$ difference in the value of $\theta$. The error on $\theta$ (as estimated from the interpolated intercept of the OCF at $\tau = 0$ \cite{LeNagard2019}) was $\epsilon = 1.0^\circ$ for the elliptical fit and $\epsilon = 2.5^\circ$ for the sine fit. The error on $\psi$ was estimated by the same method and found to be significantly larger, $\epsilon = 5.6^\circ$ on average.
The cell lengths obtained using both methods were strongly correlated, with the length measured using the elliptical fit (length of the major axis) on average $28\%$ larger than that measured with the sine fit (end-to-end distance). We decided to use the median length obtained from the sine fits as the measurement of the cell length.

\subsection{Orientation correlation functions}

The orientation correlation function (OCF) relative to the apparent orientation of the cell in the focal plane, defined as:
\begin{equation}
 C_{\perp}(\tau)=~\langle \cos \left[ \theta(t+\tau)-\theta(t) \right] \rangle, 
 \end{equation}
 was calculated for each cell by averaging over all pairs of angles separated by a given lag time $\tau$. If the cell is confined to the focal plane and in the absence of a magnetic field, the exponential form $C_{\perp} (\tau) = e^{- D_{r_{\perp}} \tau}$ is expected, with a characteristic decay time inversely related to the transversal rotational diffusion coefficient $D_{r_{\perp}}$ \cite{Nadkarni2013,Saragosti2012}. The OCF relative to the orientation of the cell around its longitudinal axis, defined as:
\begin{equation}
C_{\parallel}(\tau)=~\langle \cos \left[ \psi(t+\tau)-\psi(t) \right] \rangle,
\end{equation}
was calculated in the same way. In this case it is expected that $C_{\parallel} (\tau) = e^{- D_{r_{\parallel}} \tau}$.

\section{Results}

\subsection{\textit{M. magneticum} structural parameters}

\subsubsection{Cell body dimensions}

\begin{figure}
\includegraphics[width=8 cm]{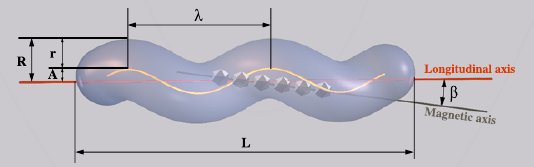}
\caption{Three-dimensional model of a \textit{M. magneticum} cell. The helical backbone, longitudinal axis (which is also the propulsion axis) and magnetic axis (assumed to be exactly aligned with the magnetosome chain) are represented by yellow, red and brown lines, respectively. }
\label{fig:geometry}
\end{figure}

The structural parameters of {\it M. magneticum} AMB-1 cells (length $L$, wavelength $\lambda$ and amplitude $A$ of the cell backbone, cell body radius $r$ and overall helical cell radius $R$, as illustrated in Fig.~\ref{fig:geometry}) were determined from phase microscopy images as explained in section~\ref{tracking}, and are summarized in Table~\ref{tab:parameters}. The length of the cells varied from $2$ to $4~\upmu$m in normal growth conditions, but increased noticeably upon addition of cephalexin (Supplementary Fig.~S1). Other cell body characteristics did not vary noticeably across the population or with growth conditions. Although the cell backbone amplitude is just below the resolution limit of the microscope ($A = 0.20~\upmu$m on average), its value is obtained with great accuracy since it comes from the fit of the position of a cloud of points. This is demonstrated by the fact that the same value is obtained for $A$ regardless of the image resolution: At $40 \times$ magnification, we measured $\lambda = 2.55 \pm 0.25 ~\upmu$m and $A = 0.20 \pm 0.04 ~\upmu$m (mean $\pm$ standard deviation (SD), $n = 44$), while at $100 \times$ magnification, we obtained $\lambda = 2.23 \pm 0.30 ~\upmu$m and $A = 0.21 \pm 0.04 ~\upmu$m (mean $\pm$ SD, $n = 31$). Values of $r$ and $R$, on the other hand, might be slightly overestimated, the first because it is obtained by direct measurement of a thickness close to the diffraction limit, and the second because it comes from an elliptical fit of a helical structure.

\subsubsection{Cell body handedness}

\begin{figure}
\includegraphics[width=8 cm]{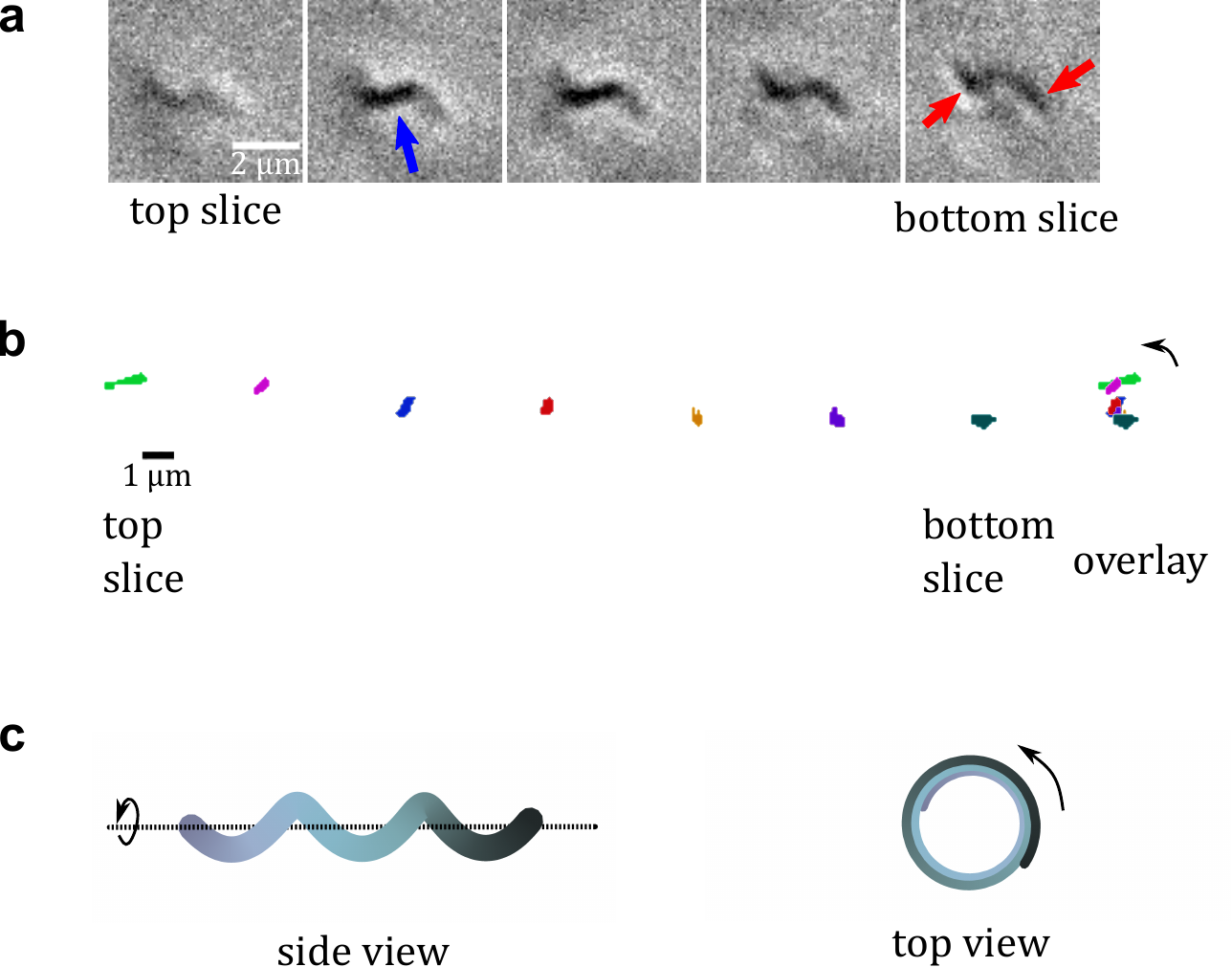}
\caption{(a): $z$-stack images (contrast-adjusted) of a cell attached to a glass slide, from top to bottom (vertical steps: $1~\upmu$m). The blue and red arrows indicates the middle and outer parts of the cell, respectively. (b) $z$-stack images of a vertical cell fixed in a gel. Colorized images are shown individually from top to bottom, and then as an overlay (vertical steps: $0.8~\upmu$m). The black arrow indicates the trace of the cell body as the cell is imaged from top to bottom. (c) Schematic side view and top view of a left-handed helix.}
\label{fig:cell_handedness}
\end{figure}

Species from the genus \textit{Aquaspirillum} all have helical cell bodies, which can be right- or left-handed \cite{konishi1986determination}. As  there had been contradictory reports concerning the handedness of \textit{M. magneticum} AMB-1 \cite{murat2015opposite,schmitzeroptical}, we set out to determine the handedness of AMB-1 cells by taking $z$-stack images of immobilized cells. Imaging the body of cells attached to the microscope coverslip from top to bottom always revealed a pattern characteristic of a left-handed helix, as illustrated in Fig.~\ref{fig:cell_handedness}a. The same was true of cells fixed in a gel and orientated perpendicular to the focal plane which, when imaged at different positions along the cell long axis, all displayed a counterclockwise pattern characteristic of left-handed helices (Fig.~\ref{fig:cell_handedness}b,c). 

The trajectories of flagellated bacteria swimming close to a solid surface also give indications about cell handedness. Hydrodynamic forces opposite in direction are exerted by the surface on a cell's rotating body and flagella, and this creates a torque on the cell resulting in a circular trajectory \cite{Lauga2009,giacche2010hydrodynamic}. We observed that cells close to a glass coverslip all had counter-clockwise trajectories when observed from the water side of the water/glass interface (see Supplementary movies). This corresponds to the motion of cells whose flagellum is rotating clockwise (when looking from the back of the cell) and is therefore right-handed \cite{Lauga2009}. The rotation of the cell body must then be in the counter-clockwise direction, presumably making it left-handed. Thus all our observations point to a left-handed cell body. 

It is intriguing that we measure a handedness that is different from that reported in Ref.~\cite{murat2015opposite}. However, we note that in that work handedness was inferred from images of horizontal cells taken in a single plane, which can give the impression that a cell has a different handedness depending on whether it lays slightly above or below the focal plane (see Fig.~\ref{fig:cell_handedness}a).

\subsection{Simulations of the rotational diffusion of an elongated magnetic particle} \label{simulations}

In order to determine the optimal experimental conditions to measure the diffusion coefficients of \textit{M. magneticum} cells, we first performed simulations of the rotational diffusion of a cell placed in an external magnetic field ($\vec{B}$), as described in section~\ref{methods:simulations}. In these simulations, the cell was represented by an elongated rigid particle (longitudinal axis $\vec{L}$) with a magnetic moment ($\vec{\mu}$) separated from $\vec{L}$ by a fixed angle ($\beta$).  

\subsubsection{Orientation distributions} \label{orientationdis}

 \begin{figure}
 \includegraphics[width=7.5 cm]{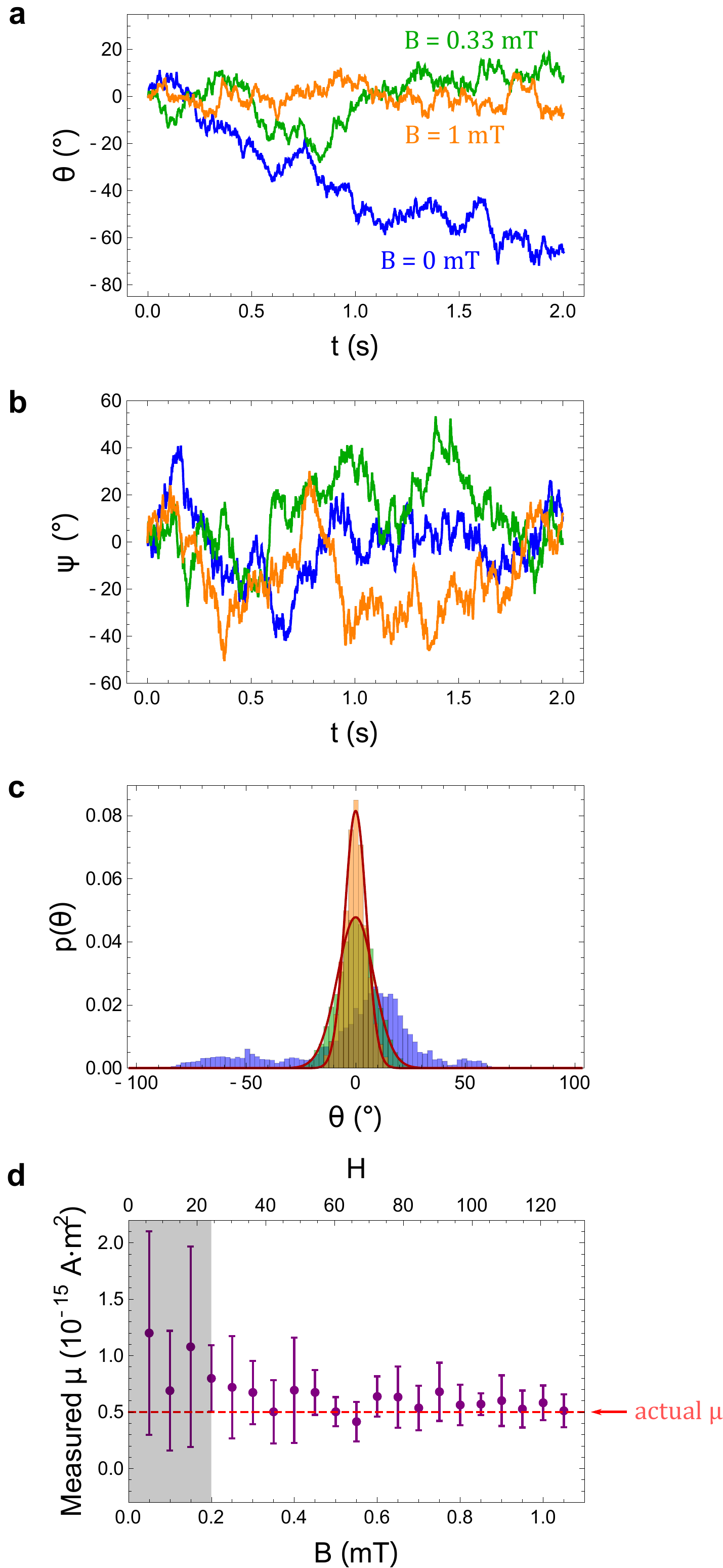}
 \caption{Angular trajectories obtained from simulations of elongated magnetic particles ($\beta = 0$). (a,b) Examples of angular trajectories for $\theta (t)$ and $\psi (t)$ at different magnetic field strengths. (c) Orientation distributions for $\theta$ obtained as a result of $20$~s simulations, with a fit to a simple Boltzmann distribution (Eq.~\ref{eq:boltzmann}). Same color scheme as in (a,b). (d) Values of the magnetic moment obtained from the fit of $\theta$ distributions (mean $\pm$ SD for n = 10 simulations each equivalent to a $2$~s experiment).}
\label{fig:fit1}
\end{figure}

We first considered the simple case where $\vec{\mu}$ is aligned with $\vec{L}$ ($\beta = 0$) and monitored the apparent orientation of the particle in the focal plane, $\theta(t)$, and around its longitudinal axis, $\psi(t)$. Examples of angular trajectories are shown in Fig.~\ref{fig:fit1}, illustrating the increased alignment of the particle with the magnetic field as $B$ increases (Fig.~\ref{fig:fit1}a). In contrast, the motion around the longitudinal axis is not affected by magnetic field strength (Fig.~\ref{fig:fit1}b). The probability distribution for $\theta$ is expected to follow a Boltzmann distribution, which has a simple form if $\vec{L}$ is restricted to the focal plane \cite{Nadkarni2013}:
\begin{equation}
p(\theta) = e^{H \cos \theta}/\left[ 2 \pi I_0(H) \right],
\label{eq:boltzmann}
\end{equation}
where $I_n$ is the modified Bessel function of the first kind of order $n$, and $H = \mu B / kT$ represents the balance between magnetic and thermal forces.
Fits of the simulated orientation distributions with Eq.~\ref{eq:boltzmann} give an estimate of $H$, from which $\mu$ can be calculated (Fig.~\ref{fig:fit1}c,d). 

Eq.~\ref{eq:boltzmann} is only strictly valid for particles constrained to rotations in the focal plane. Thus as $B$ increases and the particles pass from a free 3D to a quasi-2D motion, we expect the values of $\mu$ obtained by fitting particle orientation distributions with Eq.~\ref{eq:boltzmann} to become more accurate. Indeed our simulations show that for $B \geq 0.2$~mT, the correct value of $\mu$ is recovered (Fig.~\ref{fig:fit1}d).

\subsubsection{Orientation correlation functions}  \label{OCF}

 \begin{figure}
 \includegraphics[width=8 cm]{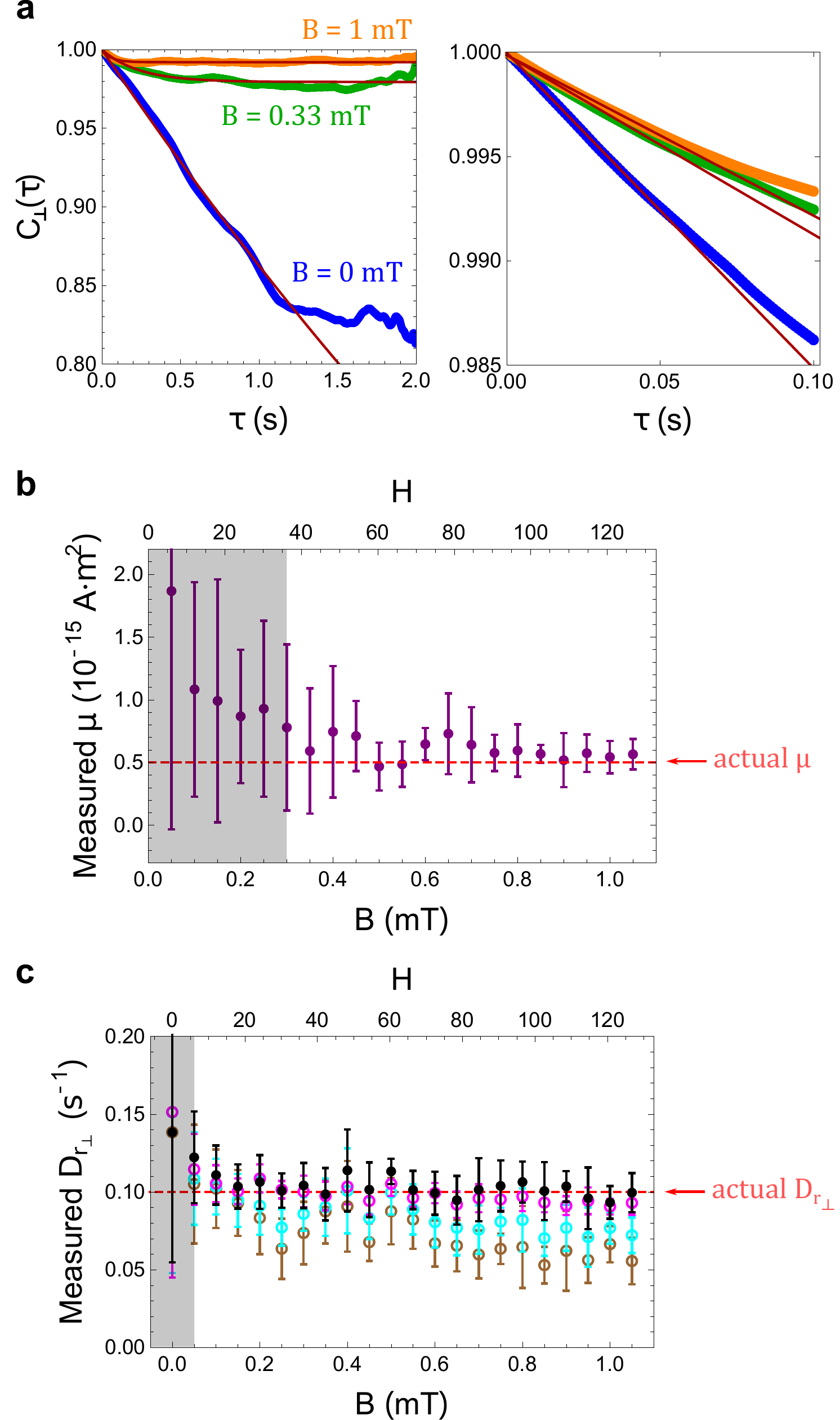}
 \caption{Simulation results for the rotation perpendicular to the particle longitudinal axis. (a) Left panel: OCF for $\theta (t)$, $C_{\perp}(\tau)$, at different magnetic field strengths, fit with Eq.~\ref{eq:OCF} for the first 1 s. Right panel: Close up on the first 100 ms of the OCF with linear fit for the first 50 ms. (b) Values of the magnetic moment $\mu$ obtained from fitting the OCF with Eq.~\ref{eq:OCF} for the first 1 s. (c) Value of the rotational diffusion coefficient $D_{r_{\perp}}$ obtained from the fit of $C_{\perp}(\tau)$. Magenta, cyan and brown empty symbols are for linear fits on the first 10 ms, 50 ms and 100 ms of the OCF, respectively. Black symbols are the result of a fit on the first 100 ms of the OCF with Eq.~\ref{eq:OCF}. In (b,c) all data points show the mean $\pm$ SD for n = 10 simulations each equivalent to a $2$~s experiment. }
\label{fig:fit2}
\end{figure}

We used our simulations to explore in which conditions the rotational diffusion coefficients of the particle could be recovered from OCF. In principle, the rotational diffusion coefficient of elongated particles perpendicular to their long axis, $D_{r_\perp}$, can be extracted from $C_{\perp}(\tau)$, the OCF related to the particle's apparent orientation in the focal plane $\theta (t)$ \cite{Saragosti2011}. When only thermal fluctuations influence the rotational diffusion and when the diffusion is restricted to the focal plane, the OCF takes an exponential form, with a characteristic decay time equal to the rotational diffusion persistence time $\tau_P = 1/D_{r_\perp}$. However, in the presence of a magnetic field, the orientation of a magnetic particle such as a MTB will become correlated at long time, and the OCF will tend towards $ I_1(H)/I_0(H)$ \cite{Nadkarni2013}. We have thus previously proposed the following empirical expression for the OCF \cite{Nadkarni2013}:
\begin{equation}
C_{\perp}(\tau) = (1- \frac{I_1(H)}{I_0(H)}) e^{-\tau/\tau'_P} + \frac{I_1(H)}{I_0(H)}.
\label{eq:OCF}
\end{equation}
At very short lag time, below $\tau_C = f_{r_\perp} / (\mu B)$, thermal motions are expected to dominate and the OCF should decay at the rate of $-D_{r_{\perp}}$ regardless of the value of $H$, thus we should have $\tau'_P = (1- I_1(H)/I_0(H)) / D_{r_\perp}$. 
The simulated OCF are indeed fitted well with Eq.~\ref{eq:OCF} (Fig.~\ref{fig:fit2}a), and the fit returns both $\mu$ and $D_{r_\perp}$, calculated from the values of $H$ and $\tau'_P$ extracted from the fit, respectively (Fig.~\ref{fig:fit2}b,c). Because Eq.~\ref{eq:OCF} was written assuming a 2D trajectory, the values obtained for small fields are not accurate, but as soon as $H \geq 5$ (i.e. $B \geq 0.05$~mT for a typical $\mu \simeq 0.5 \times 10^{-15}$~A$\cdot$ m$^2$ \textit{M. magneticum} AMB-1 cell) an accurate measurement of $D_{r_\perp}$ is obtained (Fig.~\ref{fig:fit2}c, black symbols). We also verified that $D_{r_\perp}$ could be estimated by simply fitting OCF with a linear function over a short time range (Taylor expansion of $C_{\perp}(\tau)$ at small $\tau$), which is useful when dealing with OCF calculated from short and noisy angular trajectories. We found that $D_{r_\perp}$ could indeed be determined this way as long as the linear fit is done for $\tau < 50$~ms (Fig.~\ref{fig:fit2}c, empty symbols).

\begin{figure}
 \includegraphics[width=8 cm]{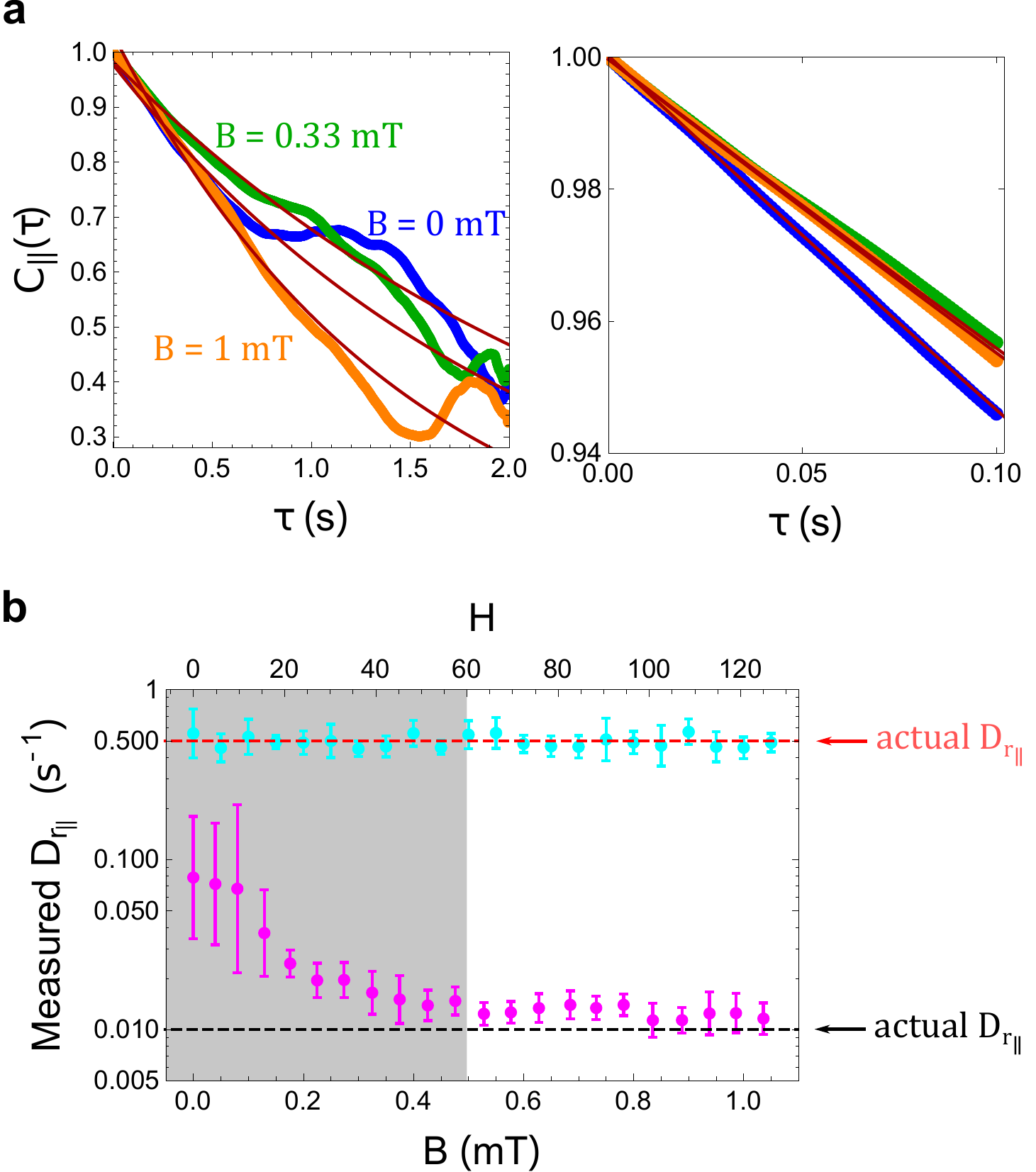}
 \caption{Simulation results for the rotation parallel to the particle longitudinal axis. (a) Left panel: OCF for the estimated $\psi (t)$, $C_{r_\parallel}(\tau)$, at different fields, with exponential fits for the first 1 s. Right panel: Close up on the first 100 ms of the OCF with exponential fit for the first 50 ms. (b) Values of the rotational diffusion coefficient $D_{r_{\parallel}}$ values obtained from fitting $C_{r_\parallel}(\tau)$. Here, magenta and cyan empty symbols are for exponential fits on the first 50 ms of the OCF (mean $\pm$ SD for n = 10 simulations each equivalent to a $2$~s experiment).}
\label{fig:fit3}
\end{figure}

The same analysis was done for the rotational diffusion of the particle around its long axis, using two different values of $D_{r_\parallel}$. This rotation is not affected by the presence of a magnetic field, thus $C_{r_\parallel}(\tau)$ is expected to exponentially decay to zero with an initial rate $-1/D_{r_\parallel}$, independently of $B$. This is what we observed when simulating particles with $D_{r_\parallel} = 0.5$~s$^{-1}$ (Fig.~\ref{fig:fit3}a), allowing an accurate measurement of $D_{r_\parallel}$ at all fields from the fit of the OCF (Fig.~\ref{fig:fit3}b). However, when using a lower $D_{r_\parallel}= 0.01$~s$^{-1}$ value in the simulations, we observed that the slope of $C_{r_\parallel}(\tau)$ changed with $B$ and that the measurements of $D_{r_\parallel}$ obtained from the fit of the $C_{r_\parallel}(\tau)$ were inaccurate at low $B$ (Fig.~\ref{fig:fit3}b). This is because the OCF is calculated from values of $\psi$ estimated as the angle between the vertical plane containing the long axis of the particle (red plane in Fig.~\ref{fig:axes}) and a vector perpendicular to $\vec{L}$. This is done in order to exactly reproduce what happens in experiments, where $\psi$ is experimentally accessible only from the analysis of the shape of the projection of the cell body in the focal plane. The estimated $\psi$ is a good approximation of the real $\psi$ only when $\vec{L}$ is close to being aligned with the focal plane. We conclude that it is safer to use $H > 60$ (i.e. for particles with a magnetic moment similar to that of a typical AMB-1 cell, $B > 0.5$~mT) in order to accurately measure $D_{r_\parallel}$ from the fit of the OCF.

\subsection{Experimental observation of \textit{M. magneticum} diffusion}

\subsubsection{Rotation perpendicular to the cell longitudinal axis}

 \begin{figure}
 \includegraphics[width=8 cm]{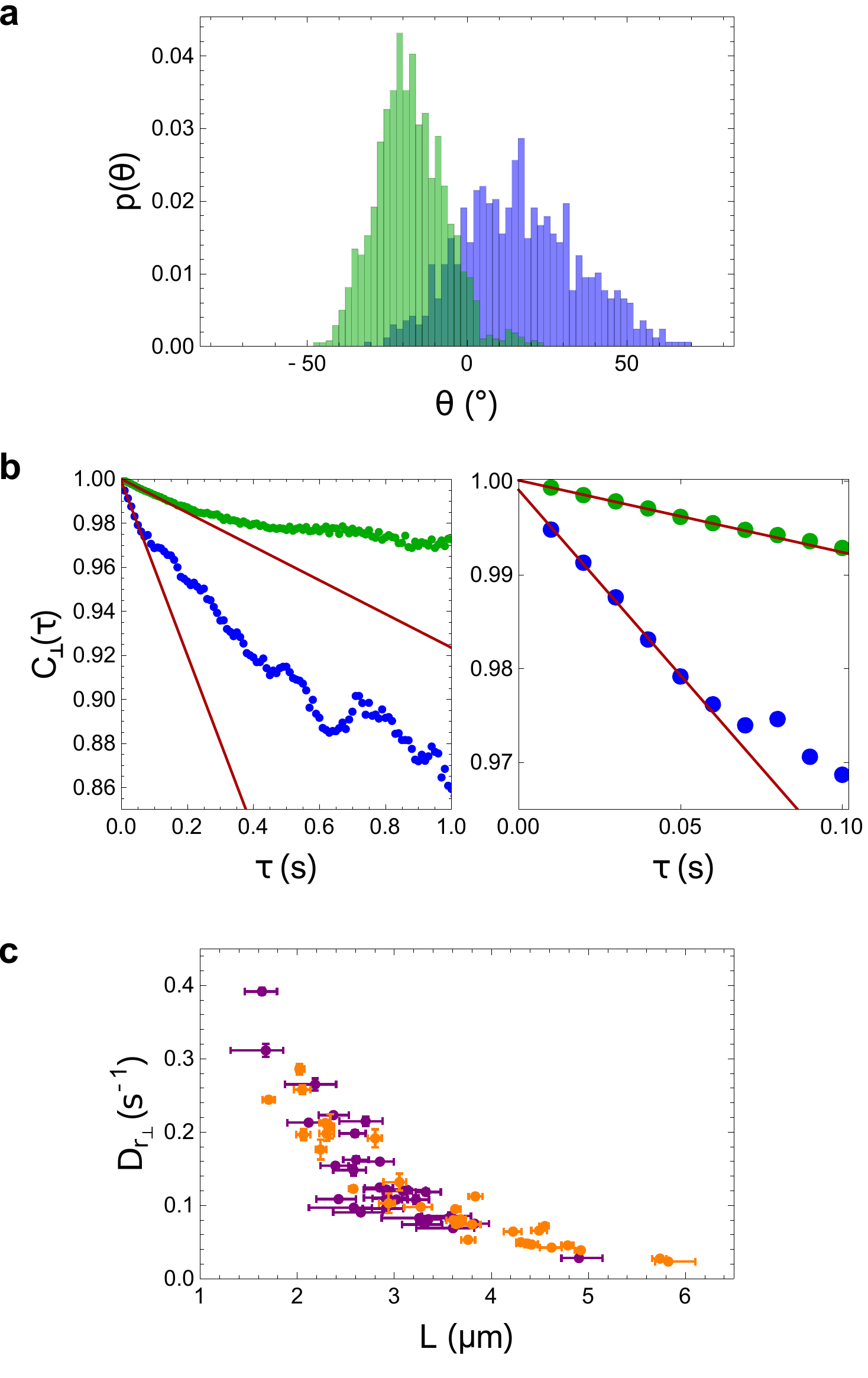}
 \caption{Experimental observation of the rotational diffusion of cells perpendicular to their longitudinal axis. (a) Examples of orientation distributions (angle $\theta$) recorded for a short cell ($L=1.6~\upmu$m, blue bars) and a long cell ($L=3.8~\upmu$m, green bars) over $8.38$ s ($B = 0.1$ mT), and (b) corresponding OCF (same color scheme as in (a), lines are linear fits based on the first 0.05 s of the OCF). The right panel is a close up on the short time range of the OCF.  
 (c) Measured rotational diffusion coefficient $D_{r_{\perp}}$ as a function of cell length (purple symbols: data obtained at $40 \times$ magnification with $B = 0.1$ mT; orange symbols: data obtained at $100 \times$ magnification with $B = 0.2$ mT, and including cephalexin treated cells). The error on $L$ was estimated using the first and third quartiles of the lengths measured for that cell over entire movie. The error on $D_{r_{\perp}}$ is the SD of values obtained from the fit of the OCF over different time ranges from $0$ to $50$~ms.} 
\label{fig:ocf1}
\end{figure}

To study the rotational diffusion of cells around an axis perpendicular to their longitudinal axis, we recorded the motion of \textit{M. magneticum} AMB-1 cells rendered non-motile by a short heat treatment. This treatment both kills and deflagellates the cells \cite{kobayashi1959purification}. Cells were then placed in low uniform magnetic fields $B = 0.1$ to $0.2$ mT, as our simulations suggested that $B = 0.1 - 0.2$~mT represented an optimal trade-off between ensuring that the rotational motion of the bacteria was detectable and yet more or less constrained to the focal plane. The orientation of the cells in the focal plane ($\theta$) and body length ($L$) were obtained by fitting the image of the cells in each available movie frame, as explained in section~\ref{tracking} and Fig.~\ref{fig:cellfit}. In normal growth conditions most cells have a length between $L = 2$ and $4~\upmu$m (as observed in previous studies \cite{LeNagard2019,Nadkarni2013}). To explore a broader range of cell lengths, we also used cells grown in the presence of $10~\upmu$g/mL cephalexin, which increased this range to $L = 3.5 - 6~\upmu$m.

Orientation distributions and OCF were generated for each cell (Fig.~\ref{fig:ocf1}a,b). The orientation distributions were usually not centered around $\theta = 0$ (Fig.~\ref{fig:ocf1}a), a sign that only the relaxation associated with rotational diffusion was observed during the finite observation time (about $10$ s), and not the relaxation associated with the rotation of the cell body around the average direction of the magnetic moment (expected for cells with a misalignment between $\vec{L}$ and $\vec{\mu}$, as discussed in section \ref{OCF} and Supplementary Fig.~S2,3). Thus a simple linear analysis of the OCF at short lag times $\tau$ was performed for each cell (Fig.~\ref{fig:ocf1}b) in order to obtain the value of its rotational diffusion coefficient $D_{r_{\perp}}$. As expected for elongated particles, $D_{r_{\perp}}$ sharply and monotonously decreases as cell length increases (Fig.~\ref{fig:ocf1}c).

\subsubsection{Rotation around the cell longitudinal axis}

 \begin{figure}
 \includegraphics[width=8 cm]{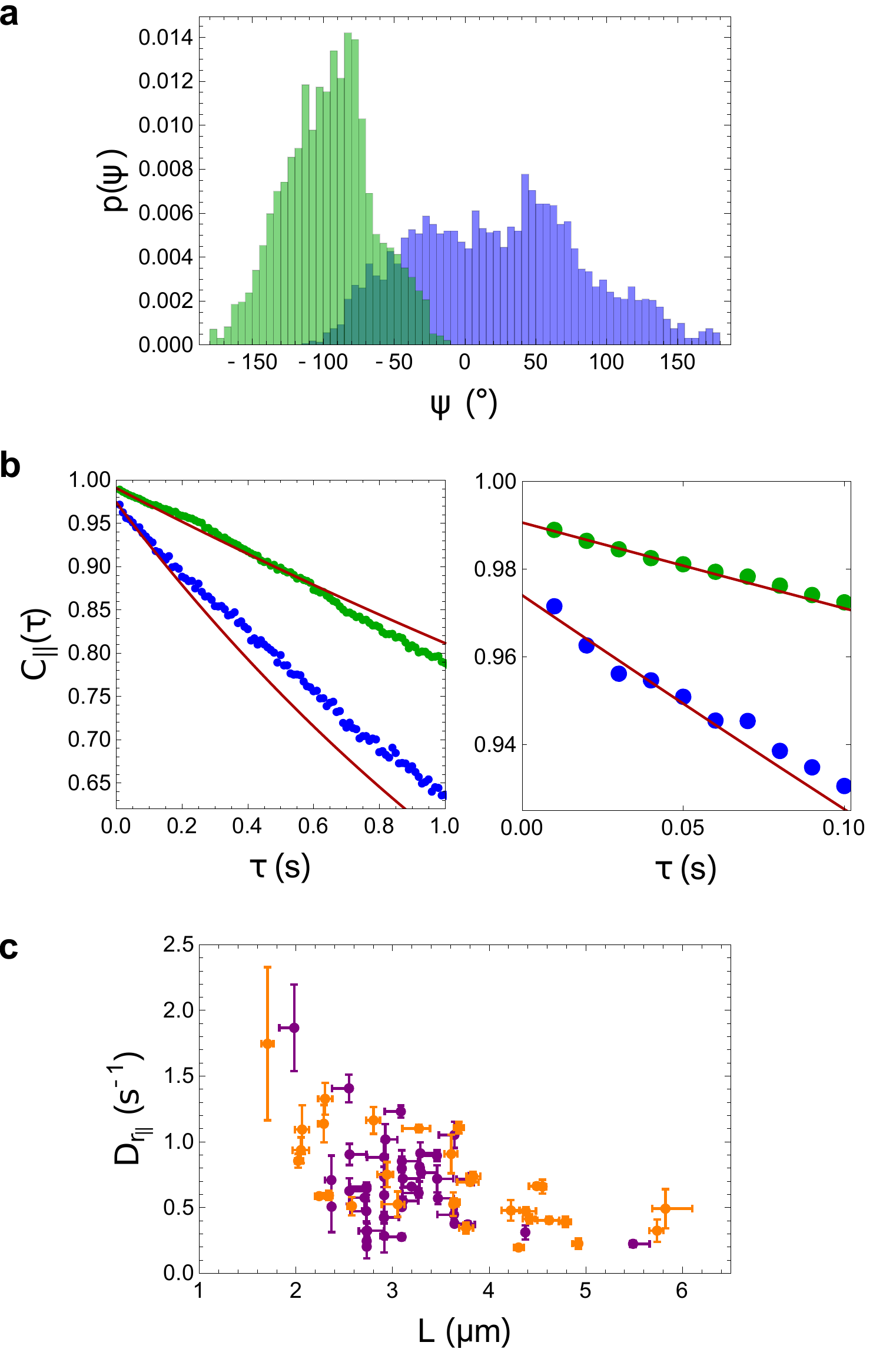}
 \caption{Experimental observation of the rotational diffusion of cells around their longitudinal axis. (a) Examples of orientation distributions (angle $\psi$) obtained for a short ($L =2.4~\upmu$m, blue bars) and a long cell ($L = 5.5~\upmu$m, green bars) over $21$ s at $B = 1$ mT. (b) Corresponding OCF (same color scheme as in (a), right panel is a close-up on the first $0.1$ s of the OCF). Red lines are exponential fits of the OCF for the first $0.05$ s. (c) Measured rotational diffusion coefficient $D_{r_{\parallel}}$ as a function of cell length. Purple and orange symbols represent data obtained at $40 \times$ and $100 \times$ magnification, respectively. Error bars were calculated as in Fig.~\ref{fig:ocf1}c.}\label{fig:ocf2}
\end{figure}

To study the rotation of the cells around their longitudinal axis, the experimental protocol was modified in two ways. First, we used a higher magnetic field ($B = 1$ mT), since our simulations suggested that $D_{r_{\parallel}}$ could be correctly estimated from the fit of the OCF only for $B > 0.5$ mT. Second, for each image, the shape of the projection of the cell backbone in the focal plane was determined and fitted to a sine function to estimate both the cell orientation in the focal plane, $\theta$, and its the angular position around its longitudinal axis, $\psi$ (see section~\ref{tracking} for details). The relationship between $\theta$ and $\psi$ allowed us to obtain, for each cell, the misalignment angle $\beta$ between magnetic moment and longitudinal axis (see Supplementary Fig. S2 and Supplementary Information for details) as done in reference \cite{LeNagard2019}. Examples of orientation distributions and OCF associated with $\psi$ are shown in Fig.~\ref{fig:ocf2}a,b. From the intercept of these OCF at $\tau=0$, it is clear that the error made on $\psi$ (Fig.~\ref{fig:ocf2}b) is much larger than the one made on $\theta$ (Fig.~\ref{fig:ocf1}b). However, this error decreases as $L$ increases (see Supplementary Fig.~S5,6). Exponential fit of the OCF associated with $\psi$ returned an estimate for the rotational diffusion coefficient $D_{r_{\parallel}}$ for each cell. Cells with a misalignment angle $\beta \geq 10^\circ$ were omitted from the results altogether, since accurate estimates of $D_{r_{\parallel}}$ then become difficult (see Supplementary Fig.~S4). Despite the  scattering in the data, it is clear that $D_{r_{\parallel}}$ decreases when the cell length increases, and that in general $D_{r_{\parallel}}$ is higher than $D_{r_{\perp}}$, as expected for an elongated particle (Fig.~\ref{fig:ocf2}c).

\subsubsection{Translational diffusion}

 \begin{figure}
	\includegraphics[width=8.5 cm]{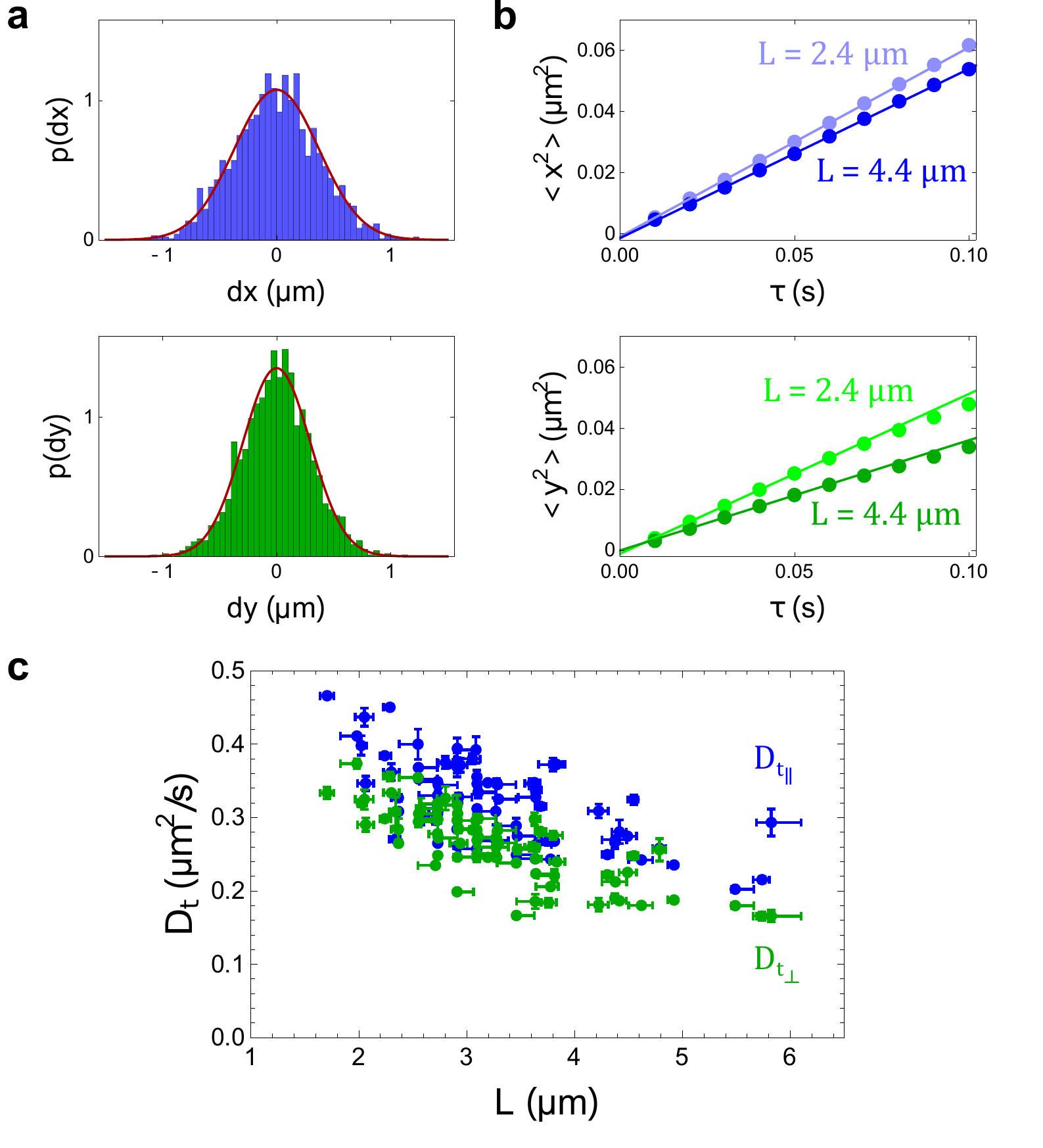}
	\caption{Experimental observation of the cells translational diffusion. (a) Examples of distributions of displacements after a time interval $\tau = 10$ ms along (upper panel, blue bars) and perpendicular (lower panel, green bars) to the cell longitudinal axis, here for a cell with length $L = 4.4~\upmu$m at $B = 1$ mT. (b) MSD and linear fit for $\tau = 0$ to 0.05 s for a short cell (light colors, $L = 2.4~\upmu$m) and a long cell (dark colors, $L = 4.4~\upmu$m). (c) Translational diffusion coefficients along ($D_{t_{\parallel}}$, blue symbols) and perpendicular ($D_{t_{\perp}}$, green symbols) to the cell long axis. Error bars were calculated as in Fig.~\ref{fig:ocf1}c.}
	\label{fig:translation}
\end{figure}

Data obtained at high magnetic field give the opportunity to estimate the two principal translational friction coefficients of the cells, since constraining their direction along that of the external magnetic field allows easily separating diffusion along and perpendicular to the cell longitudinal axis. Distributions of displacements along ($x$ - direction) and perpendicular ($y$ - direction) to the cell longitudinal axis are Gaussian (Fig.~\ref{fig:translation}a), as expected for a simple diffusion process. The mean-squared displacement (MSD) as a function of lag time was calculated for each cell in both directions (examples are shown in Fig.~\ref{fig:translation}b). Linear fits of these MSD at short lag times returned the corresponding translational diffusion coefficients. Both $D_{t_{\parallel}}$ and $D_{t_{\perp}}$ clearly decrease with cell length (Fig.~\ref{fig:translation}c), with the diffusion coefficient perpendicular to the cell longitudinal axis ($D_{t_{\perp}}$) on average smaller than the diffusion coefficient parallel to that axis ($D_{t_{\parallel}}$).

\subsubsection{Coupling between rotation and translation}

\begin{figure}
	\includegraphics[width=8 cm]{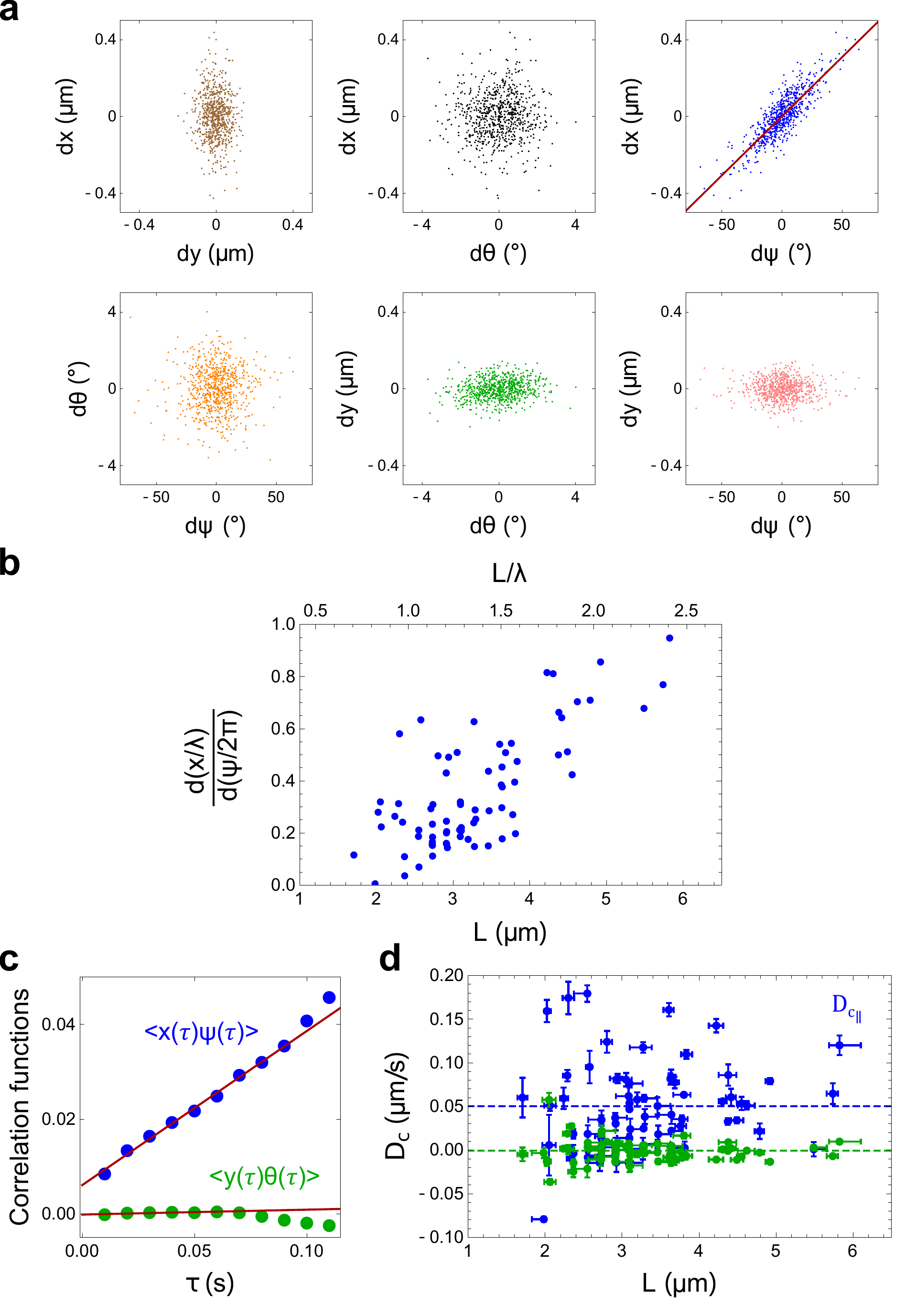}
	\caption{Experimental observation of the coupling between rotation around and translation along the cell longitudinal axis. (a) Example of the relationship observed between different types of displacements ($dx$, $dy$) and rotations ($d\theta$, $d\psi$) for one particular cell. Each point represents the cell's displacement during a $\Delta t  =0.01~$s time interval (i.e. between two consecutive frames). The red line in the $dx$ vs. $d\psi$ plot is a linear fit of the data. (b) $d x / d \psi$ in units of $\lambda/ 2 \pi$ for all studied cells. (c) Correlation functions $\langle x(\tau) \psi(\tau) \rangle$ and $\langle y (\tau) \theta(\tau) \rangle$ calculated for a particular cell, with linear fit for $\tau = 0$ to $0.05~$s. (d) Coupling diffusion coefficient parallel to the cell long axis $D_{c_{\parallel}} = \langle x(\tau) \psi(\tau) \rangle/(2t)$ (blue symbols, average of $0.05$~$\upmu$m/s) compared to $\langle y (\tau) \theta(\tau) \rangle/(2t)$ (green symbols, average of $0$~$\upmu$m/s). Error bars were calculated as in Fig.~\ref{fig:ocf1}c. }
	\label{fig:coupling}	
\end{figure}

For chiral objects such as helices, a coupling between the rotation around and translation along the helical axis is expected. We indeed detected such a coupling for individual cells, as evidenced by a correlation between the displacement along the cell longitudinal axis ($dx$) and rotation around that axis ($d\psi$) when observed between two consecutive frames (Fig.~\ref{fig:coupling}a). In contrast, no such correlation was observed for any other pairs of displacements ($dx, dy$) and rotations ($d\theta, d\psi$). The coupling between $dx$ and $d\psi$ was quantified in two ways. First, we considered the average value of $d x / d \psi$ for each cell, which we found increased linearly with cell length and approached the maximal value of $\lambda/2  \pi$ for long cells (Fig.~\ref{fig:coupling}b). Second, we looked at the correlation function $\langle x(\tau) \psi (\tau) \rangle$, which should be equal to $2 D_{c_{\parallel}} \tau$ (see Appendix). We indeed observe that $\langle x(\tau) \psi (\tau) \rangle$ is linear at short lag times (Fig.~\ref{fig:coupling}c), although the correlation is often lost at larger $\tau$. Using only the very short-term part of the correlation function, we measured the coupling diffusion coefficient along the cell long axis to be $ D_{c_{\parallel}} \simeq$ 0.05  $\upmu$m/s on average, in very clear contrast to what is observed perpendicular to the cell long axis (Fig.~\ref{fig:coupling}d).

\section{Discussion}

 \begin{figure}
 \includegraphics[width=8.5 cm]{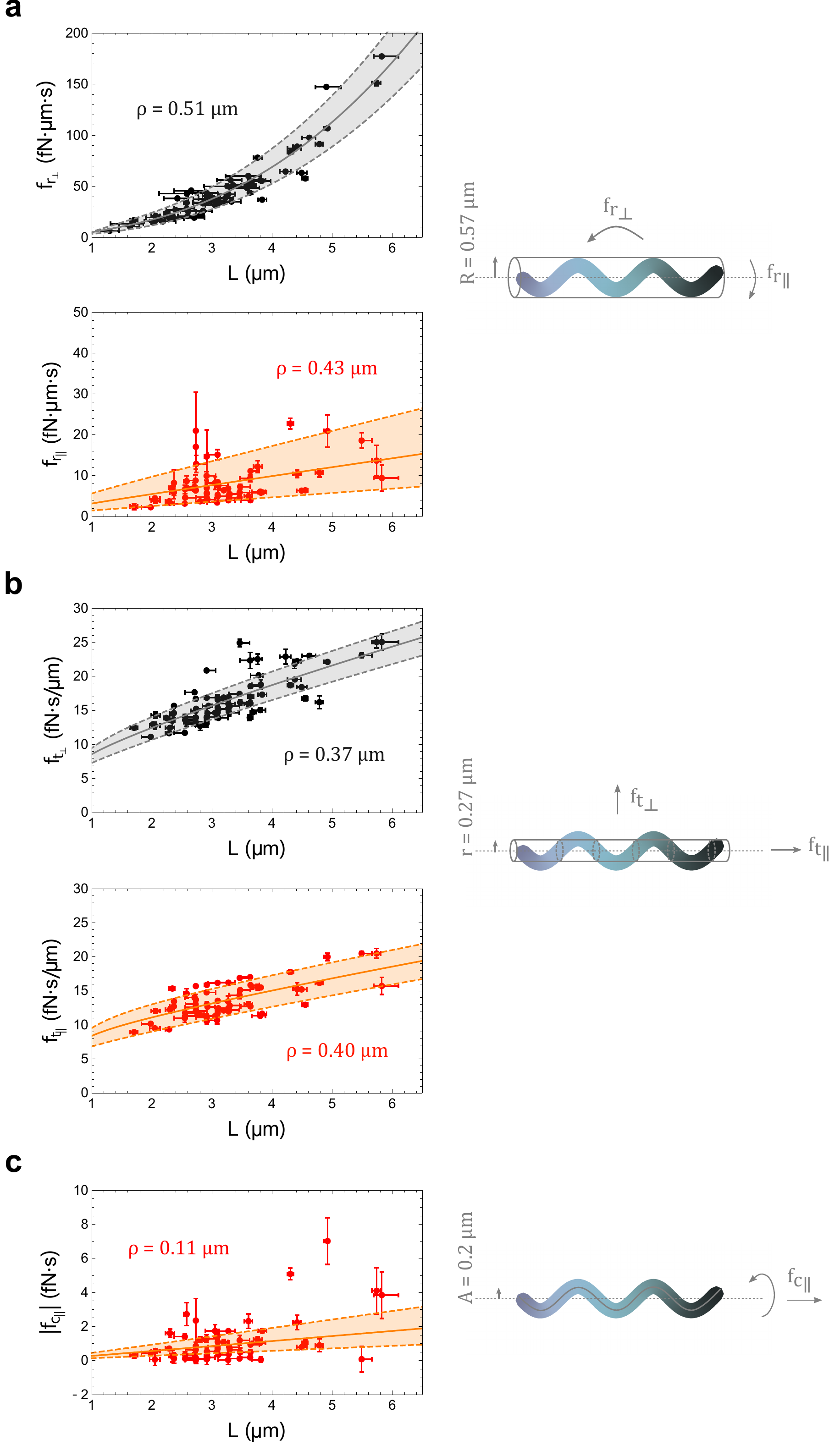}
 \caption{Dependence of AMB-1 friction coefficients on cell length. 
(a) Rotational friction coefficients, ${f_{r}}_{\perp}$ (upper panel, black symbols) and ${f_{r}}_{\parallel}$ (lower panel, red symbols). The solid lines show a fit of the binned data with the expressions expected for a cylinder of radius $\rho$ (Eqs. \ref{cylrot1} and \ref{cylrot2}), giving $\rho = 0.51~\upmu$m ( ${f_{r}}_{\perp}$) and $\rho =  0.43~\upmu$m ( ${f_{r}}_{\parallel}$), very close to the overall cell radius $R$ (as illustrated in the right panel). The shaded areas represent the range of values expected for up to $30 \%$ variations of $\rho$.  
(b) Translational friction coefficients, ${f_{t}}_{\perp}$ (upper panel, black symbols) and ${f_{t}}_{\parallel}$ (lower panel, red symbols). The solid lines are fit of the binned data with the expressions expected for a cylinder of radius $\rho$ (Eqs. \ref{cyltrans1} and \ref{cyltrans2}), giving $\rho = 0.37~\upmu$m (${f_{t}}_{\perp}$) and $\rho = 0.40~\upmu$m (${f_{t}}_{\parallel}$). These values are intermediate between $R$ and the cell body radius, $r$. The shaded areas represent the range of values expected for up to $30 \%$ variations of $\rho$.  
(c) Axial coupling friction coefficient. The solid line is a fit of the binned data with the expressions expected for a helix of radius $\rho$, yielding $\rho = 0.11~\upmu$m, very close to the value of the cell backbone amplitude, $A$, as illustrated on the right. The shaded areas the values expected for up to $30 \%$ variations of $\rho$.
}
\label{fig:f}
\end{figure}


The detection of the position and orientation of non-motile \textit{M. magneticum} AMB-1 cells allowed us to separately measure the five diffusion coefficients necessary to fully characterize their Brownian motion. From these five diffusion coefficients, the five friction coefficients found in their propulsion matrix can be calculated, fully defining the drag forces experienced by the elongated chiral cell body. For rotations and translations perpendicular to the cell longitudinal axis, for which there is no coupling, the friction coefficients ($f$) are simply related to the corresponding diffusion coefficients by $ f_{i\perp}=kT/ D_{i\perp} $, where $i = t$ or $r$. In the axial direction, however, a clear coupling between rotation and translation is detected (Fig.~\ref{fig:coupling}). Although it remains relatively weak ($ D^2_{c_{\parallel}} / (D_{t_{\parallel}} D_{r_{\parallel}}) \lesssim 7 ~ \%$ according to our measurements), it needs to be taken into account when calculating the axial translational and rotational friction coefficients (Eqs.~\ref{eq:fandD1} and \ref{eq:fandD2} in the Appendix). The last friction coefficient, the axial coupling friction coefficient, is given by: $f_{c_{\parallel}} = kT  D_{c_\parallel}/(D_{t_\parallel}D_{r_\parallel}-D^2_{c_{\parallel}})$ (Eq.~\ref{eq:fandD3}). The five friction coefficients of the body of AMB-1 cells, taking into account the correction due to coupling in the axial direction, are plotted as a function of cell body length in Fig.~\ref{fig:f}.


The friction coefficients of bacteria are often estimated by assimilating the cell body to a particle with simple geometry, usually a sphere or a cylinder. The translational and rotational friction coefficients of cylinders, which have been calculated with great precisions for a range of aspect ratios (see Appendix), can serve as a useful comparison to those of the helical \textit{M. magneticum} cells. For each friction coefficient, we binned the data by cell length, and fit the resulting curves with the expression expected for a cylinder (see Supplementary Fig.~S7). We found that both the rotational and the translational coefficients of the cells were very close to those expected for a cylinder. But whereas for rotations cells behave as cylinders of radius $\rho  = 0.43 $ to $0.51 ~\upmu$m (close to the overall radius of the cell, $R = 0.57 ~\upmu$m), for translations they behave as cylinders of radius $\rho \simeq 0.37$  to $0.40 ~\upmu$m (a value closer to the radius of the cell body, $r = 0.27~\upmu$m). Thus for rotations the overall dimensions of the cell body is what matters most, whereas for translations the exact radius of the body is also important. In all cases, the friction coefficients of \textit{M. magneticum} cells are on average larger than that of a cylinder with a radius equal to that of the cell body, a result in agreement with theoretical estimates (using Stokesian dynamics and the Boundary Integral Method) of some of the friction coefficients of another spirillum, \textit{Magnetospirillum gryphiswaldense} MSR-1  \cite{zahn2017measurement}. But in contrast to what was reported for MSR-1, we did not observe any variation of the cell overall radius with cell length and accordingly we see that on average AMB-1 friction coefficients vary with cell length exactly as expected for a cylinder of constant radius. There is, however, a lot of dispersion in the data, which cannot entirely be explained by experimental errors. Instead, invoking a $30\%$ variation in $\rho$ accounts for this dispersion (Fig.~\ref{fig:f}), which suggests that cells with the same length might have slightly different radius or morphology, maybe due to the presence of different appendages (for example because of incomplete deflagellation). 

For the non-zero axial coupling friction coefficient of AMB-1 cells, a better theoretical model for comparison with the data is that of a thin helical filament, for which $f_{c_{\parallel}}$ can be estimated (Eq.~\ref{helixrot} in the Appendix). Despite the dispersion in the data, the trend observed is in agreement with the slow increase with cell length expected for a helical filament of radius $\rho = 0.11~ \upmu$m $\pm~30\%$ (Fig.~\ref{fig:f}c and Supplementary Fig.~S7), close to the value of the cell backbone amplitude. The large dispersion in the data in this case is certainly due to the relatively large error made on the detection of the axial rotation (Fig.~\ref{fig:ocf2}).


For translational motions, we find that $f_{t_\perp} >  f_{t_\parallel}$ (Fig.~\ref{fig:f}b), as expected for elongated particles, but far from $f_{t_\perp} = 2  f_{t_\parallel}$, the expected limit for long and thin filaments \cite{hancock1953self}. Indeed, the aspect ratio for the \textit{M. magneticum} cells studied here was limited to the range $p = 1.5$ to $6$. Their dimensions are similar to that of other small spirilla such as \textit{S. gracile}, however many other spirilla are longer, with larger aspect ratios \cite{hylemon1973genus}. The thin filament approximation might thus be appropriate for other spirilla. 
For rotational motions, we also have $f_{r_\perp} >  f_{r_\parallel}$, and this is especially pronounced at large $L$ (Fig.~\ref{fig:f}a). Thus the cell body is optimized for rotations around its long axis (as happens during flagellar swimming), but not for rotations perpendicular to the cell's long axis (as may happen during changes in cell orientation). For the natural range of AMB-1 cell lengths ($2$ to $4~\upmu$m), the characteristic time scale for changes in direction of the cell axis due to rotational diffusion is $\tau_{r_\perp} = 1/D_{r_\perp}=3$ to $30$ s. This is too slow for rapid changes in swimming direction, and indeed AMB-1 cells do not make use of rotational diffusion when they need to change direction. Instead, since they are amphitrichous, they reverse their propulsion direction by changing the direction of rotation their flagella \cite{murat2015opposite}. Other bacteria, such as E. coli, solve this problem in a different way, by using active rotational diffusion during tumbles in order to speed up changes in direction \cite{Saragosti2012}.

\begin{center}
\begin{table*}
\begin{tabular} { | M{4cm} M{2cm} M{2cm} M{2cm} M{2cm} M{2cm} M{2.4cm} | }
\hline
 $L$ ($\upmu$m) & $\lambda$  ($\upmu$m) &  $A$  ($\upmu$m) & $r$   ($\upmu$m) & $R$  ($\upmu$m) & $\beta$ ($^{\circ}$)  & Handedness  \\
\hline
2 - 4 (no cephalexin) &  \multirow{2}{2cm}{$2.4 \pm 0.3 $}  &  \multirow{2}{2cm}{$0.20 \pm 0.04$}  &  \multirow{2}{2cm}{$ 0.27 \pm 0.07 $} &  \multirow{2}{2cm}{$ 0. 57 \pm 0.08$} &  \multirow{2}{2cm}{$6.5 \pm 3.2$}  & \multirow{2}{2.4cm}{Left-handed} \\ 
3.5 - 6 (with cephalexin) &   &  &  &  &   &  \\ 
\hline
\end{tabular}
\caption{Physical parameters characterizing the geometry of the body of {\it M. magneticum} AMB-1 cells. All reported values were experimentally measured in this study ($L$, $\lambda$ and $A$ from sine fits of $n = 75$ cells, $R$ from ellipsoidal fit of $n = 31$ cells imaged at $100 \times$ magnification, $r$ from $n=10$ cells imaged at $100 \times$ magnification) except for $\beta$ whose value was taken from Ref.~\cite{LeNagard2019}. Average values are reported as mean $\pm$ SD deviation, except for $r$ where the error corresponds to the size of a single pixel in the image used to measure this parameter.}
\label{tab:parameters}
\end{table*}
\end{center}


The measurement of the axial friction coefficients allows to calculate the drag force $\cal{F}^{\text{body}}_{\text{drag}}$ and torque $\cal{L}^{\text{body}}_{\text{drag}}$ applied to the AMB-1 cell body when swimming (average swimming speed $V \simeq 20 ~\upmu$m/s and average angular velocity $\Omega \simeq 200~$rad/s). Since $f_{t_\parallel} = 13 \pm 2 ~$fN$\cdot$s$/\upmu$m and $f_{r_\parallel} = 7 \pm 4 ~$fN$\cdot$$\upmu$m$\cdot$s (mean $\pm$ SD for $2.5 - 3.5~\upmu$m long cells) we find that $\cal{F}^{\text{body}}_{\text{drag}}$ $= f_{t_\parallel}  V  \simeq 0.26 \pm 0.04~$pN and $\cal{L}^{\text{body}}_{\text{drag}}$ $ = f_{r_\parallel}  \Omega \simeq 1 \pm 1 ~$pN$\cdot$$\upmu$m. The drag force on the cell body is likely much larger than that on the flagella, so the overall drag force on the cell is $\cal{F}^{\text{cell}}_{\text{drag}} \simeq \cal{F}^{\text{body}}_{\text{drag}} $. Because swimming takes place as low Reynold's number, $\cal{F}^{\text{cell}}_{\text{drag}}$ is also equal to the total propulsive force (thrust) of the cell, which is therefore on the order of $ \simeq 0.26 \pm 0.04~$pN. This value is comparable to the thrust estimated for \textit{E. coli} and \textit{Salmonella typhimurium} \cite{chattopadhyay2006swimming,hughes1999measurement}, but significantly larger than that previously estimated for AMB-1 \cite{pierce2019thrust}.

Further considering that $|f_{c_{\parallel}}| \simeq$ 0.7 $\pm$ 0.7 fN$\cdot$s allows us to calculate the propulsive thrust due to the rotation of the cell body alone: $\cal{F}^{\text{body}}_{\text{thrust}} $$ = |f_{c_{\parallel}}| \Omega \simeq$ 0.1 $\pm$ 0.1 pN. It is interesting that propulsive thrust is comparable to the drag force $\cal{F}^{\text{body}}_{\text{drag}}$ experienced by the cell body, because it suggests that the chiral shape of the body of AMB-1 cells is an important contribution to the cell propulsion, significantly adding to the propulsion contributed by the flagella. A different conclusion was reached for \textit{Helicobacter pylori}, from hydrodynamic calculations based on body and flagella shape and dimensions \cite{constantino2016helical}. However, the cell body diameter of \textit{M. magneticum} is thinner than that of \textit{H. pylori}.


Friction coefficients had never, to our knowledge, been measured directly for any type of bacterial cell before this study. Our results illustrate the fact that slight differences in dimensions can results in large differences in friction coefficients, especially rotational friction coefficients. This highlights the importance of single cell characterization for precise studies of bacterial swimming motions, or for studies where friction needs to be precisely estimated in order to measure propulsion, magnetic or optical torques using torque balance (e.g. measurements of the torque generated by the flagellar motor \cite{lowe1987rapid,berg1993torque,chen2000torque} or measurement of the magnetic moment of a cell with the U-turn method \cite{Esquivel1986}).

\begin{acknowledgments}
This work was funded by the Natural Sciences and Engineering Research Council of Canada (NSERC).
\end{acknowledgments}


\section{Appendix: Friction coefficients of an elongated particle} \label{appendix}

\subsection{Propulsion matrix}

The drag forces on a rigid body are characterized by the friction coefficient tensor (also known as resistance matrix or propulsion matrix):
\begin{equation}
\cal{K} = 
\begin{pmatrix} 
A & B \\ 
B^T & D 
\end{pmatrix}
.
\end{equation}
At low Reynold's numbers, this tensor can be used to express the external force and torque, $\vec{\cal{F}}$ and $\vec{\cal{L}}$, applied to the object, as a function of its velocity and angular velocity, $\vec{V}$ and $\vec{\Omega}$:
\begin{equation}
\begin{pmatrix} 
\vec{\cal{F}} \\ 
\vec{\cal{L}}
\end{pmatrix}
= \cal{K} 
\begin{pmatrix} 
\vec{V} \\ 
\vec{\Omega}
\end{pmatrix}
.
\end{equation}

\subsection{Translation matrix for a short cylinder}

For a particle with revolution symmetry, and chosing the $x$-axis aligned with the symmetry axis, the translation submatrix is diagonal:
\begin{equation}
A = 
\begin{pmatrix} 
f_{t_\parallel} & 0 & 0 \\
0 & f_{t_\perp} & 0  \\
0 & 0 & f_{t_\perp}
\end{pmatrix}
,
\end{equation}
where $f_{t,\parallel}$  and $f_{t,\perp}$ are the translational drag coefficients parallel and perpendicular to the object long axis. 

For a sphere of diameter $L$, $f_{t_{\parallel}} = f_{t_{\perp}} = 3 \pi \eta L$ (where $\eta$ is the solvent viscosity). But for an elongated particle (length $L$, radius $\rho$), $f_{t_{\parallel}}<f_{t_{\perp}}$, and both coefficients vary with the aspect ratio of the particle, $p = L/(2\rho)$. For cylinders with $2<p<30$, these coefficients were calculated with great precision by modelling the particle surface with a series of beads and found to be well approximated by \cite{tirado1979translational,de1981hydrodynamic,tirado1984comparison}:
\begin{equation}
f_{t_{\parallel}} \simeq \frac{ 2 \pi \eta L}{\ln p - 0.207+0.980/p-0.133/p^2} ,  
\label{cyltrans1}
\end{equation}
and:
\begin{equation}
f_{t_{\perp}} \simeq \frac{ 4 \pi \eta L}{\ln p + 0.839+0.185/p+0.233/p^2}.
\label{cyltrans2}
\end{equation}

\subsection{Rotation matrix for a short cylinder}

The rotation submatrix of a particle with revolution symmetry is also diagonal:
\begin{equation}
D = 
\begin{pmatrix} 
f_{r_\parallel} & 0 & 0 \\
0 & f_{r_\perp} & 0  \\
0 & 0 & f_{r_\perp}
\end{pmatrix}
.
\end{equation}

For a sphere, $f_{r_{\parallel}} = f_{r_{\perp}} = \pi \eta L^3$, while for cylinders with $2<p<30$ a good approximation is \cite{tirado1980rotational,de1981hydrodynamic}: 
\begin{equation}
f_{r_\parallel}  \simeq  \pi \eta L R^2 \times 3.84 \left[1+0.677/p-0.183/p^{2} \right] 
\label{cylrot1}
\end{equation} 
for axial rotations, and:
\begin{equation}
f_{r_\perp} \simeq \frac{ \pi \eta L^3} {3 \left[ \ln p-0.662+0.917/p-0.050/p^2 \right] }
\label{cylrot2}
\end{equation}
for rotations about the cylinder short axes. 

\subsection{Coupling matrix for a thin helix}

The coupling matrix is $B = 0$ for a particle with true revolution symmetry, meaning that rotations are decoupled from translational motions \cite{perrin1936mouvement, han2006brownian}. However, for a chiral particle such as a helix there is a coupling between axial translation and rotation, and therefore:
\begin{equation}
B = 
\begin{pmatrix} 
f_{c_\parallel} & 0 & 0 \\
0 & 0 & 0  \\
0 & 0 & 0
\end{pmatrix}
,
\end{equation}
where $f_{c_\parallel} > 0$ for a right-handed helix and $f_{c_\parallel} < 0 $ for a left-handed helix. 

For a thin left-handed helix with length $L$, radius $\rho$, pitch $\lambda$ and $\rho \ll \lambda$, one can show that \cite{Lauga2009}:
\begin{equation}
f_{c_\parallel} \simeq  -  \left(  f_{r_\parallel} - \rho^2 f_{t_{\parallel}}  \right) \frac{2 \pi}{\lambda}.
\label{helixrot}
\end{equation} 

\subsection{Relationship between friction coefficients and diffusion coefficients}

For a particle undergoing Brownian motion, writing the Langevin equations and applying the equipartition theorem leads to the diffusion tensor $\cal{D}$ = $kT$ $\cal{K}$$^{-1}$, and to the following expressions for the mean-squared displacements and rotations of the object \cite{Hoshikawa1979}:
\begin{equation}
\begin{aligned}
\langle x^2(t) \rangle = 2 D_{t_\parallel} t = 2 \frac{kT f_{r_\parallel}}{f_{t_\parallel} f_{r_\parallel} - f_{c_\parallel}^2} t  \\
\langle y^2(t) \rangle =  \langle z^2(t) \rangle = 2 D_{t_\perp} t = 2 \frac{kT}{f_{t_\perp} } t \\
\langle \omega_x^2(t) \rangle =  2 D_{r_\parallel} t = 2 \frac{kT f_{t_\parallel}}{f_{t_\parallel} f_{r_\parallel} - f_{c_\parallel}^2} t  \\
\langle \omega_y^2(t) \rangle = \langle \omega_z^2(t) \rangle =  2 D_{r_\perp} t = 2 \frac{kT}{f_{r_\perp} } t  
\end{aligned}
\end{equation}
as well as the following correlations:
\begin{equation}
\begin{aligned}
& \langle x(t) \omega_x(t) \rangle  = - 2 D_{c_\parallel} t = - 2 \frac{kT f_{c_\parallel}}{f_{t_\parallel} f_{r_\parallel} - f_{c_\parallel}^2} t  \\
& \langle x(t) y(t) \rangle  = \langle \omega_x(t) \omega_y(t) \rangle = 0 \\
& \langle x(t) \omega_y(t) \rangle  = \langle \omega_x(t) y(t) \rangle =  0 \\
\end{aligned}
\end{equation}

Thus the particle friction coefficients are related to its diffusion coefficients by:
\begin{equation}
\begin{aligned}
f_{t_{\parallel}} & = kT \frac{D_{r_\parallel}}{D_{t_\parallel}D_{r_\parallel}-D_{c_\parallel}^2} \\
f_{t_{\perp}} & =  kT \frac{1}{D_{t_{\perp}}}
\label{eq:fandD1}
\end{aligned}
\end{equation}

\begin{equation}
\begin{aligned}
f_{r_{\parallel}} & = kT \frac{D_{t_\parallel}}{D_{t_\parallel}D_{r_\parallel}-D_{c_\parallel}^2} \\
f_{r_{\perp}} & =  kT \frac{1}{D_{r_{\perp}}}
\label{eq:fandD2}
\end{aligned}
\end{equation}

and:

\begin{equation}
\begin{aligned}
f_{c_{\parallel}} & = kT \frac{D_{c_\parallel}}{D_{t_\parallel}D_{r_\parallel}-D_{c_\parallel}^2}
\label{eq:fandD3}
\end{aligned}
\end{equation}


\bibliography{MTB_Biblio}

\end{document}